\documentstyle[]{article}

\input{epsf}

\renewcommand{\abstract}[1]{{ \footnotesize \noindent {\bf Abstract} #1 \\}}
\renewcommand{\author}[1]{\subsubsection*{\bf#1}}
\newcommand{\address}[1]{\subsubsection*{\it#1}}

\begin{document}

\section*{\LARGE{\bf{The effect of unbound stars on the mass modelling of the Fornax dwarf}}}

\vspace{0.3in}

\author{E. L. {\L}okas, J. Klimentowski and R. Wojtak}

\address{Nicolaus Copernicus Astronomical Center, Bartycka 18, 00-716 Warsaw, Poland}

\abstract{We discuss how different approaches to selecting member stars in kinematic samples
of dwarf spheroidal galaxies affect the estimates of their mass and anisotropy of stellar orbits.
We demonstrate
that the selection of members is an additional source of error compared to the usual uncertainties
due to the sampling of velocity moments. As an example we use the kinematic data set for 202 stars
in the Fornax dwarf galaxy for which we model the velocity dispersion profile and estimate the
mass-to-light ratio and anisotropy assuming that mass follows light. We also show that stronger
constraints on these parameters can be obtained if kurtosis of the velocity distribution
is included in the analysis. Using the Besancon model of the Milky Way we demonstrate that the majority
of contamination in Fornax probably comes from the Milky Way stars.}

\section{Which stars are members?}

Dwarf spheroidal (dSph) galaxies of the Local Group
are believed to be the most dark matter dominated galaxies. One of the
major problems in determining their masses and mass-to-light ratios lies in the selection of
members which we can trust to trace the dynamics of the galaxies reliably. Due to their
proximity to the Milky Way (MW) the galaxies are strongly affected by tidal forces, their stars are
stripped and form tidal tails. Most of the stars in the tails are not bound to the dwarf
and including them in kinematic modelling may overestimate the inferred mass-to-light ratios
dramatically. Another possible source of contamination are the stellar populations of the
MW which can however be dealt with using detailed photometric studies (Majewski et al. 2000).
On the other hand,
the tidally stripped stars will have the same photometric properties as the population
of the dwarf. It is therefore essential to reject them via other methods.
\begin{figure}[h]
\begin{center}
    \leavevmode
    \epsfxsize=10cm
    \epsfbox[100 30 430 340]{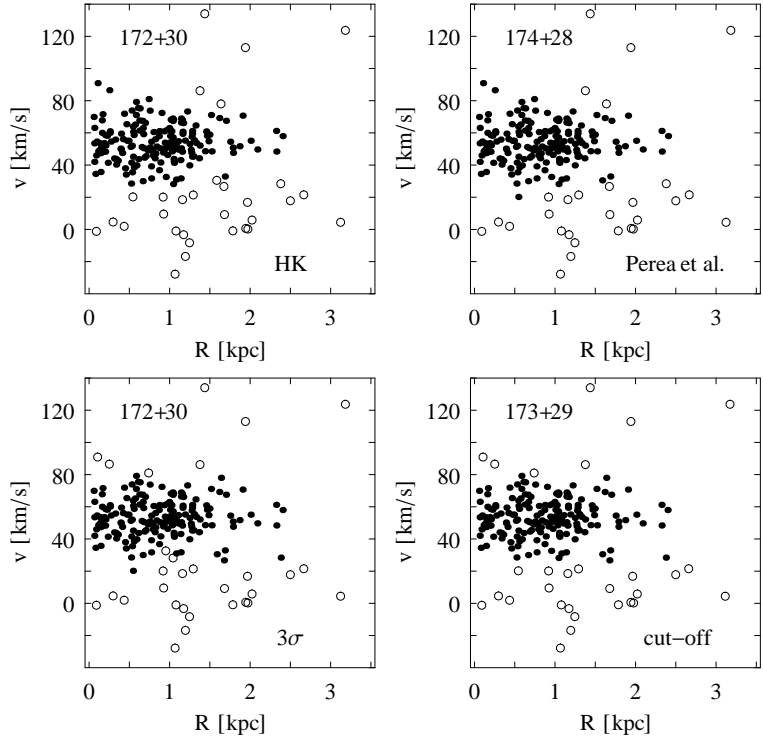}
\end{center}
\caption{The kinematic sample of 202 stars in the Fornax dwarf after application of
different schemes for the removal of unbound stars (open circles).}
\label{velocitydiagrams}
\end{figure}

We illustrate a number of methods to deal with such {\em interlopers} using a recently published
kinematic sample of 202 stars in the Fornax dwarf (Walker et al. 2006). The sample may contain
both tidally stripped stars and those from the MW since the selection has been done only by
choosing stars along the red giant branch and no colour-colour information was used.
Figure~\ref{velocitydiagrams} shows the line-of-sight
heliocentric velocities of the stars as a function of their projected distance from the dwarf
centre. In each panel the filled circles correspond to the stars adopted as members and open
circles are interlopers. We present the results of four interloper rejection methods described in
detail in Wojtak et al. (2007): the method based on finding the maximum velocity available
to a star proposed by den Hartog \& Katgert (1996) (HK), the method based on the ratio of
two mass estimators proposed by Perea et al. (1990), the method based on iterative rejection
of stars outside $\pm 3 \sigma(R)$ (where $\sigma(R)$ is the velocity dispersion in different
radial bins) and a simple constant cut-off in velocity
which retains a similar number of stars as the other methods. In the upper left corner of each
panel we give the number of members and rejected stars.
\begin{figure}[h]
\begin{center}
    \leavevmode
    \epsfxsize=10cm
    \epsfbox[100 20 430 345]{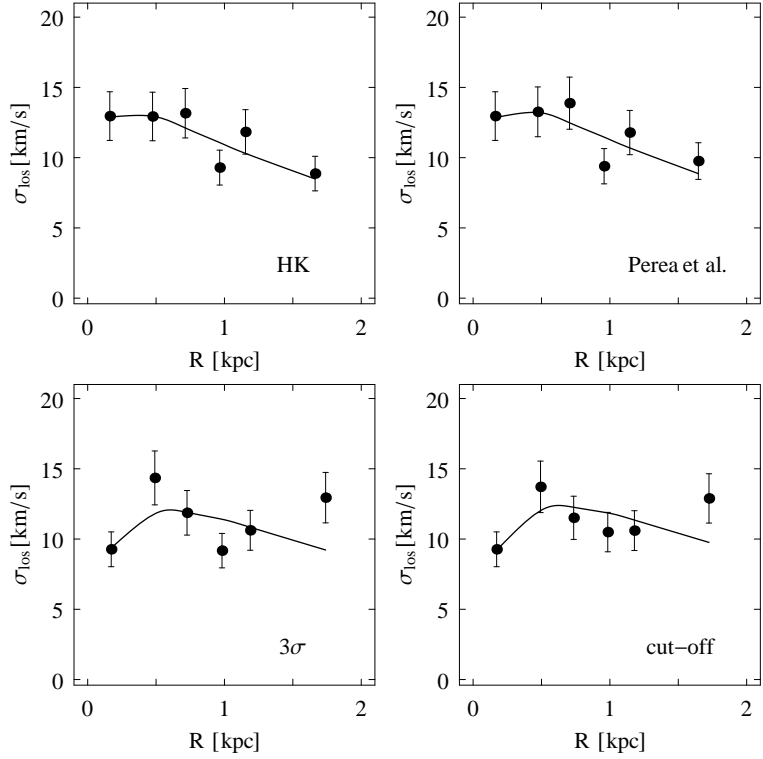}
\end{center}
\caption{Velocity dispersion profiles for the different selections of member stars in
Figure~\ref{velocitydiagrams}.}
\label{fornaxdisp}
\end{figure}

\section{Modelling of velocity moments}

\begin{figure}[h]
\begin{center}
    \leavevmode
    \epsfxsize=10cm
    \epsfbox[100 25 430 340]{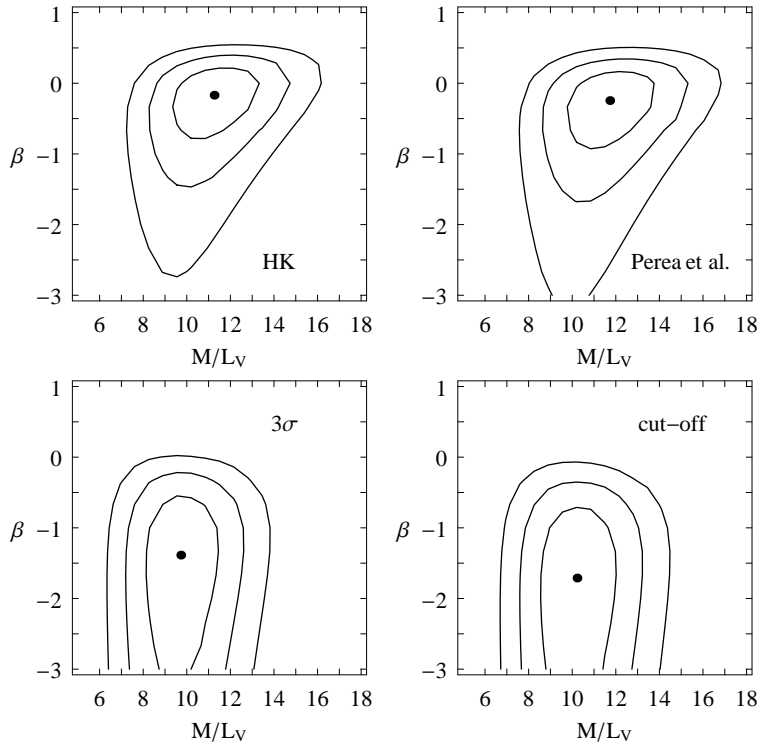}
\end{center}
\caption{The $1\sigma$, $2\sigma$ and $3\sigma$
constraints on the parameters $\beta$ and $M/L_V$ from fitting the velocity dispersion profiles
of Figure~\ref{fornaxdisp} with the assumption that mass follows light and $\beta=$const.
Dots mark the best fits.}
\label{fornaxcont}
\end{figure}

The velocity dispersion profiles calculated from the different samples are shown in
Figure~\ref{fornaxdisp}. As we can see, in the two lower panels the velocity dispersion
profiles show a secondary increase at larger $R$ signifying incomplete removal of unbound stars.
The velocity dispersion profiles have been fitted with the solutions of the Jeans
equation (see {\L}okas 2002, {\L}okas et al. 2005) assuming that mass follows light.
The parameters of the light distribution were adopted from Irwin \& Hatzidimitriou (1995).
For each sample we estimated two constant parameters, the mass-to-light ratio $M/L_V$ and
the anisotropy parameter $\beta$ which describes the type of orbits the stars follow.
The results in terms of the best-fitting parameters (dots) and $1\sigma$, $2\sigma$ and $3\sigma$
constraints on the parameters (contours) are presented in Figure~\ref{fornaxcont}. As we can see,
going from the upper left to the lower right panel, more and more negative $\beta$ values
(corresponding to more tangential orbits) are preferred. This is the consequence of the
increasing velocity dispersion profiles. If the assumption of mass following light
were relaxed this type of behaviour could also lead us to overestimate the mass (since
extended mass distribution can also fit the increasing $\sigma_{\rm los}$ profile).

The method of HK has been shown by Wojtak et al (2007) to be the most effective in removing
unbound particles from samples generated from simulated dark matter haloes.
It has also been shown to perform well
on the stellar samples drawn from simulated two-component models of dSph galaxies by
Klimentowski et al. (2007). We can therefore conclude that the results obtained for the HK
sample are the most reliable. One must however keep in mind when modelling the
kinematics of dSph galaxies that in addition to the
uncertainties due to sampling errors of velocity moments there is always additional
uncertainty due to the method by which the member stars were selected. Since the
simple cut-off in velocity is still the most common method used for interloper rejection
many results are biased towards tangential orbits or extended mass distributions.
For example, Walker et al. (2006) using the same data concluded that if mass follows light in Fornax then
isotropic orbits ($\beta=0$) are excluded while we show that the data are perfectly consistent with
$\beta=0$ if the unbound stars are properly removed from the sample.

\begin{figure}
\begin{center}
    \leavevmode
    \epsfxsize=10cm
    \epsfbox[100 15 430 170]{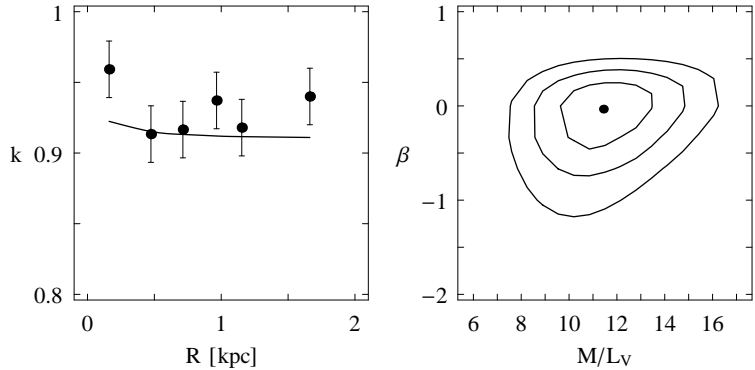}
\end{center}
\caption{The kurtosis of the line-of-sight velocity distribution in Fornax (left panel)
and the $1\sigma$, $2\sigma$ and $3\sigma$ confidence regions on the fitted parameters
from the joint fitting of dispersion and kurtosis (right panel).}
\label{fornaxkurt}
\end{figure}

Once the interlopers are reliably removed one can also consider with confidence the higher
moments of the velocity distribution. Figure~\ref{fornaxkurt} (left panel) shows the profile
of the variable $k=(\log \kappa_{\rm los})^{1/10}$ where $\kappa_{\rm los} (R)
= \overline{v_{\rm los}^4} (R)/\sigma_{\rm los}^4 (R)$ is the kurtosis of the line-of-sight
velocity distribution. The kurtosis was calculated for the most reliable sample generated by
the HK method of interloper rejection. The fourth velocity moment $\overline{v_{\rm los}^4} (R)$ is
governed by the higher-order Jeans equation (see {\L}okas 2002, {\L}okas et al. 2005).
The solid line in the Figure is the best-fitting solution when the dispersion and kurtosis
are fitted simultaneously. In the right panel we plot the $1\sigma$, $2\sigma$ and $3\sigma$
confidence regions on the fitted parameters. Comparing with Figure~\ref{fornaxcont} we see that
the anisotropy parameter is now much more constrained, at 68\% confidence we get $\beta =
-0.03^{+0.23}_{-0.37}$, $M/L_V =11.4^{+2.0}_{-1.7}$ M$_{\odot}/L_{\odot}$ with $\chi^2/N=10.9/10$
while when fitting only the dispersion profile (for the same HK sample) we had $\beta =
-0.17^{+0.37}_{-0.63}$, $M/L_V =11.3^{+2.1}_{-1.8}$ M$_{\odot}/L_{\odot}$ with $\chi^2/N=3.4/4$.
We see that when kurtosis is added the best fitting anisotropy is even closer to $\beta=0$.

\begin{figure}
\begin{center}
    \leavevmode
    \epsfxsize=6cm
    \epsfbox[0 0 370 370]{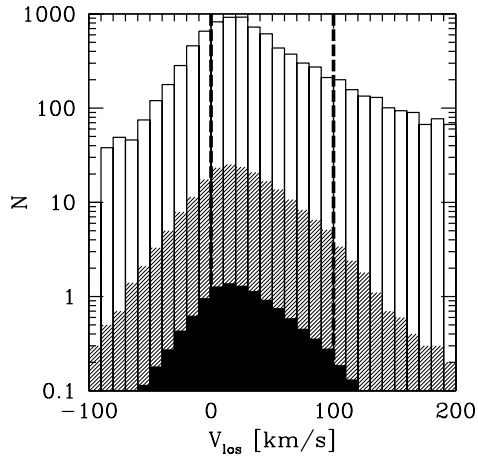}
\end{center}
\caption{Expected contamination of the Fornax kinematic sample by MW stars.}
\label{besancon}
\end{figure}

\section{The origin of unbound stars}

As discussed in detail by Klimentowski et al. (2007) using an N-body simulation of a
dSph galaxy, the contribution from the tidally stripped stars to the kinematic samples
depends strongly on the orientation of the tidal tails with respect to the line of sight.
If the tails are along the line of sight then the contamination is strongest. This kind
of contamination should also be symmetric in velocity: we should see similar amount of
unbound stars at velocities above and below the mean velocity of the dwarf galaxy.
As Figure~\ref{velocitydiagrams} demonstrates, this is not the case: there are much more
stars rejected with velocities below the systemic velocity of Fornax. There is also
evidence for the visibility of tidal tails in Fornax from the photometric study of Coleman et al.
(2005) which means that they cannot be oriented exactly along the line of sight. These two
facts lead to the conclusion that the majority of the contaminating stars in the sample come
from the MW stellar population.

We have verified this conjecture by referring to the Besancon model of the MW
(Robin et al. 2003). The results are illustrated in Figure~\ref{besancon} where we plot
the histograms showing the expected number of stars with a given velocity. The open
histogram refers to all stars expected to be seen in the direction of Fornax, the half-filled
one to stars along the red giant branch of Fornax and the filled one is for stars expected to
be seen in a small kinematic sample of 202 stars. As expected, the maximum of the distribution
falls around the velocity of 20 km s$^{-1}$ which confirms that most of the contamination
in Fornax indeed comes from the MW stars.

\end{document}